\documentclass[journal]{IEEEtran}
\ifCLASSINFOpdf
\else
  \usepackage[dvips]{graphicx}
 \DeclareGraphicsExtensions{.eps}
\fi
\usepackage[cmex10]{amsmath}
\usepackage{relsize}
\usepackage{bm}

\begin{document}
%
\title{Coupled-wave surface-impedance analysis of extraordinary transmission through single and stacked metallic screens}
%
%
%

\author{Vicente~Delgado,~\IEEEmembership{}
        Ricardo~Marqu\'es,~\IEEEmembership{}
        and~Lukas~Jelinek~\IEEEmembership{}
\thanks{V. Delgado and R. Marqu\'es are with Department of Electronics and Electromagnetism, University of Seville, 41012-Sevilla, Spain
(e-mail: vdelgado@us.es; marques@us.es).}
\thanks{L. Jelinek is with Department of Electromagnetic Field, Czech Technical University in Prague, 16627-Prague, Czech Republic
(e-mail: jelinel1@fel.cvut.cz).}
}

\maketitle

\begin{abstract}
In this paper we present an efficient Coupled-wave surface-impedance method for the analysis of extraordinary optical transmission (EOT) through single and stacked realistic metallic screens under normal and oblique incidence, including possible dielectric interlayers. The proposed theory is valid for the complete frequency range where EOT has been reported, including microwaves and optics. Electromagnetic simulations validate the results of the model, which allows for a fast and accurate characterization of the analyzed structures.
\end{abstract}


%
\IEEEpeerreviewmaketitle

\section{Introduction}
%
%
%
%
\IEEEPARstart{T}{he} phenomenon of Extraordinary Optical Transmission (EOT) through metallic screens with a periodic array of subwavelength holes, first reported in \cite{Ebbesen-1998}, still is a topic of intensive research. The reader is referred to the excellent reviews by C. Genet et al. \cite{Genet-2007}, F. J. Garc\'ia de Abajo \cite{Abajo-2007} and F. J. Garc\'ia Vidal et al. \cite{Garcia-Vidal-2010} in order to have a complete overview of the topic. Coupling of surface plasmons through holes provided the first explanation for EOT \cite{Ghaemi-1998,Pendry-2004}. However, EOT has also been reported in cases where surface plasmons cannot exist, such as metals at microwave frequencies \cite{Beruete-2004} and waveguide structures \cite{Medina-2009}. These phenomena show that EOT is a quite general effect that can be seen from many perspectives; plasmonics  \cite{Ghaemi-1998,Grupp-2000}, diffraction models \cite{Moreno-2001,Lezec-2004,Abajo-2005,Bravo-2006}, general surface-wave (or spoof-plasmon) analysis \cite{Pendry-2004,Abajo-2005b}, and circuit analysis \cite{Medina-2008,Beruete-2011}; an approach that has been promisingly extended to plasmonic structures \cite{Staffaroni-2012}. Recently, some of the authors of this paper presented an analytical theory of EOT through perfect conducting screens, based on waveguide analysis \cite{Marques-2009}. This theory was later extended to screens made of realistic conductors and metals at optical frequencies \cite{Delgado-2010} making use of the surface impedance concept, which is widely used in classical electromagnetism for analysing the skin effect in imperfect conductors \cite{Jackson}. These theories provide a different and valid perspective on EOT but, due to the inherent simplicity of the models, they still have strong limitations: they are only valid for normal incidence and for holes smaller than - approximately - a quarter wavelength. These limitations arise from the approximations made in the model, mainly from the ``small hole approximation'' first suggested by Gordon in \cite{Gordon-2007}.

EOT may find direct applications in photonic circuits \cite{Devaux-2003} and optical sensing \cite{Yanik-2010}. EOT also is closely related to fishnet metamaterials \cite{Beruete-2006-Opex,Beruete-2007-AP,Marques-2009b,Jelinek-2010}. Complete fishnet analysis requires repeated computation of many stacked periodically perforated screens excited at oblique incidence \cite{Beruete-2010-NJP,Jelinek-2010}. Therefore, a fast and accurate method for the analysis of a single EOT screen is an unavoidable step for the complete characterization of fishnet metamaterials. In \cite{Jelinek-2010} a mode-matching analysis of stacked perfect conducting EOT screens was shown to provide the necessary fast and accurate tool for this computation. However, direct application of this method to realistic metallic screens is cumbersome, due to the non-negligible penetration of fields inside metals at optical frequencies. In order to conveniently take into account this penetration without a substantial increase in complexity, the surface impedance concept offers an useful alternative \cite{Delgado-2010}.

Coupled-wave method \cite{Katsenelenbaum}, in general applicable also to non-uniform lines \cite{Johnson-2002}, is considerably simplified in our case that only exhibits piecewise variation of cross-section. In such cases, the fields on the structure are simply expanded in the eigenmodes of the infinite waveguides matching the cross section at each point, except inside the metal, where we use approximate surface-impedance boundary conditions \cite{Tretyakov,Delgado-2010} in order to relate the electromagnetic fields on the interfaces. For the specific case of screens made of perfect conductors, the method reduces to standard mode-matching analysis \cite{Collin} with ${\bf n}\times{\bf E}=0$ boundary conditions on the metallic surfaces.

Using scattering matrix formalism we will further extend the method to the case of stacked screens including dielectric substrates. This will allow us to study interesting phenomena such as anomalous EOT \cite{Kuznetsov-2009} or ``negative refraction'' of Gaussian beams \cite{Beruete-2009,Beruete-2010-NJP}.

\section{Theory} \label{Theory}
\begin{figure}[!t]
\centering
\includegraphics[width=1.0\columnwidth]{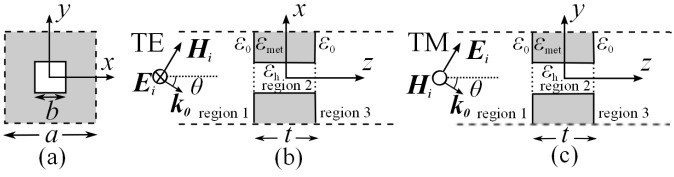}
\caption{Front (a) and side views (b, c) of the unit cell of the structure with the three regions in which the fields are expanded. The incident waves in the cases of TE and TM polarization are shown in (b) and (c), respectively. Due to the polarization of the incident waves and due to periodicity, in (b) there are virtual electric walls in $y=0,\pm a/2$ and in (c) there are virtual magnetic walls in $x=0,\pm a/2$. The rest of lateral boundaries are periodic boundaries.}
\label{Fig1}
\end{figure}

For simplicity, we will first consider a metallic screen with a periodic 2D array of square holes (the extension to rectangular holes and periodicities is straightforward). The geometry of the unit cell is shown in Fig.\ref{Fig1}.
Oblique incidence of TE and TM waves, with tangential $E$ or $H$ fields polarized along one of the main axes of the structure will be considered. Both cases are summarized in Fig.\ref{Fig1}(b) and Fig.\ref{Fig1}(c). The transversal components of the electromagnetic fields at both sides of the screen (regions 1 and 3) and inside the hole (region 2) are expanded in terms of Bloch and waveguide modes, and evaluated at the input and output surfaces of the screen (see Eqs.(\ref{12}-\ref{15}) in the Appendix). The next step is to determine the coefficients in (\ref{12}-\ref{15}). For this purpose we impose a surface impedance condition on the metallic interfaces at $z=\pm t/2$
\begin{equation}\label{01}
\begin{gathered}
  \left[ {\begin{array}{*{20}c}
   {\bm{E_{\rm{\parallel}}^{(1)}}} (z=-t/2)   \\
   {\bm{E_{\rm{\parallel}}^{(3)}}} (z=t/2)   \\

 \end{array} } \right]
  \approx \bm{\overline{\overline Z}}  \left[ {\begin{array}{*{20}c}
   \bm{\hat{z}} \times {\bm{H_{\rm{\parallel}}^{(1)}}} (z=-t/2)   \\
   \bm{\hat{z}} \times {\bm{H_{\rm{\parallel}}^{(3)}}} (z=t/2)   \\

 \end{array} } \right]  \hfill \\
  {\rm{\:for\:}} b/2<|x|,|y|<a/2\rm,
 \end{gathered}
\end{equation}

\noindent where $\bm{E_{\rm{\parallel}}^{(1)}}$, $\bm{E_{\rm{\parallel}}^{(3)}}$, $\bm{H_{\rm{\parallel}}^{(1)}}$ and $\bm{H_{\rm{\parallel}}^{(3)}}$ are vectors containing the transversal components of the fields in regions 1 and 3, and $\bm{\overline{\overline Z}}$ is the surface impedance matrix corresponding to an infinite slab (without holes) for a given incident plane wave \cite{Tretyakov}. When applying (\ref{01}) we are neglecting the perturbation associated to the presence of the holes. Another assumption is that all the refracted modes can be connected by approximately the same surface impedance, so that (\ref{01}) is satisfied by the total fields. This last approximation is valid as long as the transverse wavenumbers of the different modes ($k_{x,n}^{(1)}$, $k_{y,m}^{(1)}$) in (\ref{12}) and (\ref{14})) are much smaller than the TEM wavenumber inside the metal. This approximation will fail near the plasma frequency, which happens only for very small periodicities or for very big angles of incidence (see Section III.A).

Matrix $\bm{\overline{\overline Z}}$ is symmetric, and can be diagonalized, so that the following combinations of the tangential components of the fields vanish at the screen interfaces \cite{Delgado-2010}
\begin{equation}\label{02}
\begin{gathered}
   E_y^{(3)} (z = t/2  ) + E_y^{(1)} (z = -t/2  ) \hfill \\
   - Z_{{\rm s}1} \left[ {H_x^{(3)} (z = t/2  ) - H_x^{(1)} (z = -t/2  )} \right] \approx 0 \hfill \\
   E_y^{(3)} (z = t/2  ) - E_y^{(1)} (z = -t/2  ) \hfill \\
   - Z_{{\rm s}2} \left[ {H_x^{(3)} (z = t/2  ) + H_x^{(1)} (z = -t/2  )} \right] \approx 0 \hfill \\
   E_x^{(3)} (z = t/2  ) + E_x^{(1)} (z = -t/2  ) \hfill \\
   + Z_{{\rm s}1} \left[ {H_y^{(3)} (z = t/2  ) - H_y^{(1)} (z = -t/2  )} \right] \approx 0 \hfill \\
   E_x^{(3)} (z = t/2  ) - E_x^{(1)} (z = -t/2  ) \hfill \\
   + Z_{{\rm s}2} \left[ {H_y^{(3)} (z = t/2  ) + H_y^{(1)} (z = -t/2  )} \right] \approx 0 \hfill \\
  {\rm{\hspace{1.6cm}\:for\:}} b/2<|x|,|y|<a/2\rm, \hfill \\
\end{gathered}
\end{equation}
where $Z_{{\rm s}1}$ and $Z_{{\rm s}1}$ are the eigenvalues of $\bm{\overline{\overline Z}}$
\begin{equation}\label{03}
\begin{gathered}
  Z_{{\rm s}1}  = \frac{{1 + \cos ( {k_{z,{\rm ref}}^{({\rm met})} t} )}}
{{i\sin ( {k_{z,{\rm ref}}^{({\rm met})} t} )Y_{{\rm ref}}^{({\rm met})} }} \hfill \rm and\\
  Z_{{\rm s}2}  = \frac{{i\sin ( {k_{z,{\rm ref}}^{({\rm met})} t} )}}
{{\left[ {1 + \cos ( {k_{z,{\rm ref}}^{({\rm met})} t} )} \right]Y_{{\rm ref}}^{({\rm met})} }} \rm. \hfill \\
\end{gathered}
\end{equation}
In (\ref{03}), $k_{z,{\rm ref}}^{({\rm met})}$ is the longitudinal component (i.e. parallel to the $z$ axis) of the propagation constant of the ``refracted wave'' inside the metal, and $Y_{{\rm ref}}^{({\rm met})}$ is the wave admittance of this ``refracted wave''. In this context, the ``refracted wave'' is defined as the wave that would be refracted at the air-metal interface if there were no holes in the screen, and the wave admittance is defined as the ratio between the transverse (in the $x-y$ plane) components of the $H$ and $E$ fields of this refracted wave. For an incident TE wave as depicted in Fig. \ref{Fig1}(b)
\begin{equation}\label{031}
  k_{z,{\rm ref}}^{({\rm met})}  = \sqrt {k^2  - k_{x,0}^2 }  {\rm{\hspace{0.2cm}}} {\rm{and}}  {\rm{\hspace{0.2cm}}}   Y_{{\rm ref}}^{({\rm met})}  = \frac{k_{z,{\rm ref}}^{({\rm met})}}{k}Y^{({\rm met})},
\end{equation}\label{031}
and for an incident TM wave as depicted in Fig. \ref{Fig1}(c)
\begin{equation}\label{032}
  k_{z,{\rm ref}}^{({\rm met})}  = \sqrt {k^2  - k_{y,0}^2 }  {\rm{\hspace{0.2cm}}} {\rm{and}}  {\rm{\hspace{0.2cm}}}   Y_{{\rm ref}}^{({\rm met})}  = \frac{k}{k_{z,{\rm ref}}^{({\rm met})}}Y^{({\rm met})},
\end{equation}
where $k = \omega \sqrt {\varepsilon _{\rm{met}} \mu _0 }$ and $Y^{({\rm met})}  = \sqrt {\varepsilon _{\rm{met}} /\mu _0 }$ are the propagation constant and the intrinsic admittance of the metal with relative permittivity $\varepsilon _{\rm{met}}$.

In the area of the holes, continuity of transverse electromagnetic fields is imposed,
\begin{equation}\label{04}
\begin{gathered}
  \left[ {\begin{array}{*{20}c}
   \bm {E_{\parallel}^{(1)}} (z = -t/2)   \\
   \bm {H_{\parallel}^{(1)}} (z = -t/2)   \\

 \end{array} } \right] = \left[ {\begin{array}{*{20}c}
   \bm {E_{\parallel}^{(2)} } (z = -t/2)   \\
   \bm {H_{\parallel}^{(2)} } (z = -t/2) \\

 \end{array} } \right]  \hfill \\
  \:\:\: \left[ {\begin{array}{*{20}c}
   \bm {E_{\parallel}^{(2)}}  (z = t/2)  \\
   \bm {H_{\parallel}^{(2)}} (z = t/2)  \\

 \end{array} } \right] = \left[ {\begin{array}{*{20}c}
   \bm {E_{{\rm \parallel}}^{(3)}} (z = t/2)   \\
   \bm {H_{{\rm \parallel}}^{(3)}} (z = t/2)    \\

 \end{array} } \right] \rm  \hfill \\
 {\rm{\:\hspace{2.5cm} for\:}}|x|,|y|<b/2. \hfill \\
\end{gathered}
\end{equation}

Equations (\ref{02}) and (\ref{04}) provide a complete set of boundary conditions. However, they are not appropriate for the direct computation of the coefficients in (\ref{12}-\ref{15}). It is better to define \cite{Note}
\begin{equation}\label{05}\begin{gathered}
  \left[ {\begin{array}{*{20}c}
   \bm {E_{\parallel}^{(2)} } (z = \mp t/2)   \\
   \bm {H_{\parallel}^{(2)} } (z = \mp t/2) \\
 \end{array} } \right] = 0 \\
 {\rm{\:\hspace{0.0cm} for\:}}b/2<|x|,|y|<a/2\rm , \hfill \\
\end{gathered}
\end{equation}
and combine (\ref{02}) with (\ref{04}-\ref{05}) to obtain
\begin{equation}\label{06}
\begin{gathered}
   E_y^{(3)} (z = t/2  ) + E_y^{(1)} (z = -t/2  ) \hfill \\
   - Z_{{\rm s}1} \left[ {H_x^{(3)} (z = t/2  ) - H_x^{(1)} (z = -t/2  )} \right] \approx \hfill \\
   E_y^{(2)} (z = t/2  ) + E_y^{(2)} (z = -t/2  ) \hfill \\
   - Z_{{\rm s}1} \left[ {H_x^{(2)} (z = t/2  ) - H_x^{(2)} (z = -t/2  )} \right] \hfill \\\\
   E_y^{(3)} (z = t/2  ) - E_y^{(1)} (z = - t/2  ) \hfill \\
   - Z_{{\rm s}2} \left[ {H_x^{(3)} (z = t/2  ) + H_x^{(1)} (z = - t/2  )} \right] \approx \hfill \\
   E_y^{(2)} (z = t/2  ) - E_y^{(2)} (z = - t/2  ) \hfill \\
   - Z_{{\rm s}2} \left[ {H_x^{(2)} (z = t/2  ) + H_x^{(2)} (z = - t/2  )} \right] \hfill \\\\
   E_x^{(3)} (z = t/2  ) + E_x^{(1)} (z =-t/2  ) \hfill \\
   + Z_{{\rm s}1} \left[ {H_y^{(3)} (z = t/2  ) - H_y^{(1)} (z = -t/2  )} \right] \approx \hfill \\
   E_x^{(2)} (z = t/2  ) + E_x^{(2)} (z =-t/2  ) \hfill \\
   + Z_{{\rm s}1} \left[ {H_y^{(2)} (z = t/2  ) - H_y^{(2)} (z = -t/2  )} \right] \hfill \\\\
   E_x^{(3)} (z = t/2  ) - E_x^{(1)} (z = -t/2  ) \hfill \\
   + Z_{{\rm s}2} \left[ {H_y^{(3)} (z = t/2  ) + H_y^{(1)} (z = -t/2  )} \right] \approx \hfill \\
  E_x^{(2)} (z = t/2  ) - E_x^{(2)} (z = -t/2  ) \hfill \\
   + Z_{{\rm s}2} \left[ {H_y^{(2)} (z = t/2  ) + H_y^{(2)} (z = -t/2  )} \right] \hfill \\\\
  {\rm{\hspace{1.6cm}\:for\:}} 0<|x|,|y|<a/2\rm. \hfill \\
\end{gathered}
\end{equation}

These equations provide a complete set of four boundary conditions in the range $b/2<|x|,|y|<a/2$. We still need four more linearly independent equations in order to have a complete set of boundary conditions in the region $|x|,|y|<b/2$. For this purpose we choose
\begin{equation}\label{07}
\begin{gathered}
   \bm {H_{\parallel}^{(1)}} (z = -t/2) = \bm {H_{\parallel}^{(2)} } (z = -t/2) \\
   \bm {H_{\parallel}^{(2)}} (z = t/2) = \bm {H_{{\rm \parallel}}^{(3)}} (z = t/2) \\
 {\rm{\:\hspace{2.0cm} for\:}}|x|,|y|<b/2. \hfill \\
\end{gathered}
\end{equation}

Equations (\ref{06}) and (\ref{07}) are a complete set of boundary conditions, which provide a sparse system of equations for the mode coefficients (see Appendix). The specific case of perfect conducting screens is obtained by simply taking $ Z_{{\rm s}1}= Z_{{\rm s}2}=0$ instead of (\ref{03}) .

\section{Results}

\subsection{Single EOT screens}

\begin{figure}[!t]
\centering
\includegraphics[height=4.6cm]{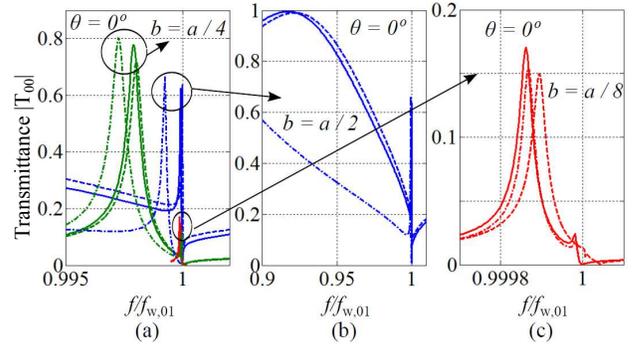}
\caption{Transmission through an array of square holes in a copper screen ($\sigma = 59.6 \times 10^6$ S/m) at normal incidence and different sizes of the holes. Periodicity is $a=300{\rm{\mu}}$m and thickness of the screen is $t=a/20$. Wood's anomaly frequency is $f_{{\rm w},01}\approx 1$ THz. Continuous lines correspond to the mode matching model, dashed lines to the CST simulations and dotted-dashed lines to our previous model \cite{Delgado-2010}. In (a) a general picture is shown; in (b) and (c) the details of the peaks for $b=a/2$ and $b=a/8$ are shown.}
\label{Fig2}
\end{figure}

\begin{figure}[!t]
\centering
\includegraphics[height=4.6cm]{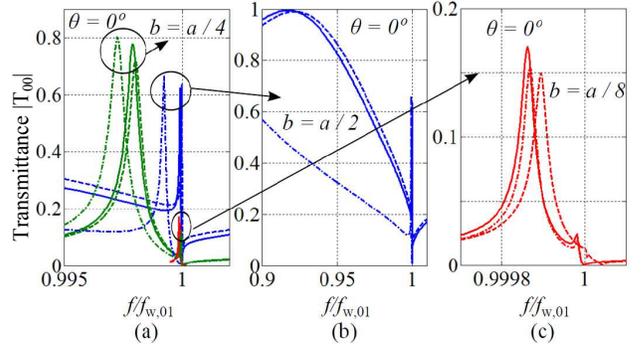}
\caption{Transmission through an array of square holes in a silver screen ($\omega_p = 2\pi\times 2175$ THz and $f_c= 4.35$ THz) at normal incidence and different sizes of the holes. Periodicity is $a=1{\rm{\mu}}$m and thickness of the screen is $t=a/20$. Wood's anomaly frequency is $f_{{\rm w},01}\approx 299.79$ THz. Continuous lines correspond to the mode matching model, dashed lines to the CST simulations and dotted-dashed lines to our previous model \cite{Delgado-2010}.}
\label{Fig3}
\end{figure}

\begin{figure}[!t]
\centering
\includegraphics[height=4.6cm]{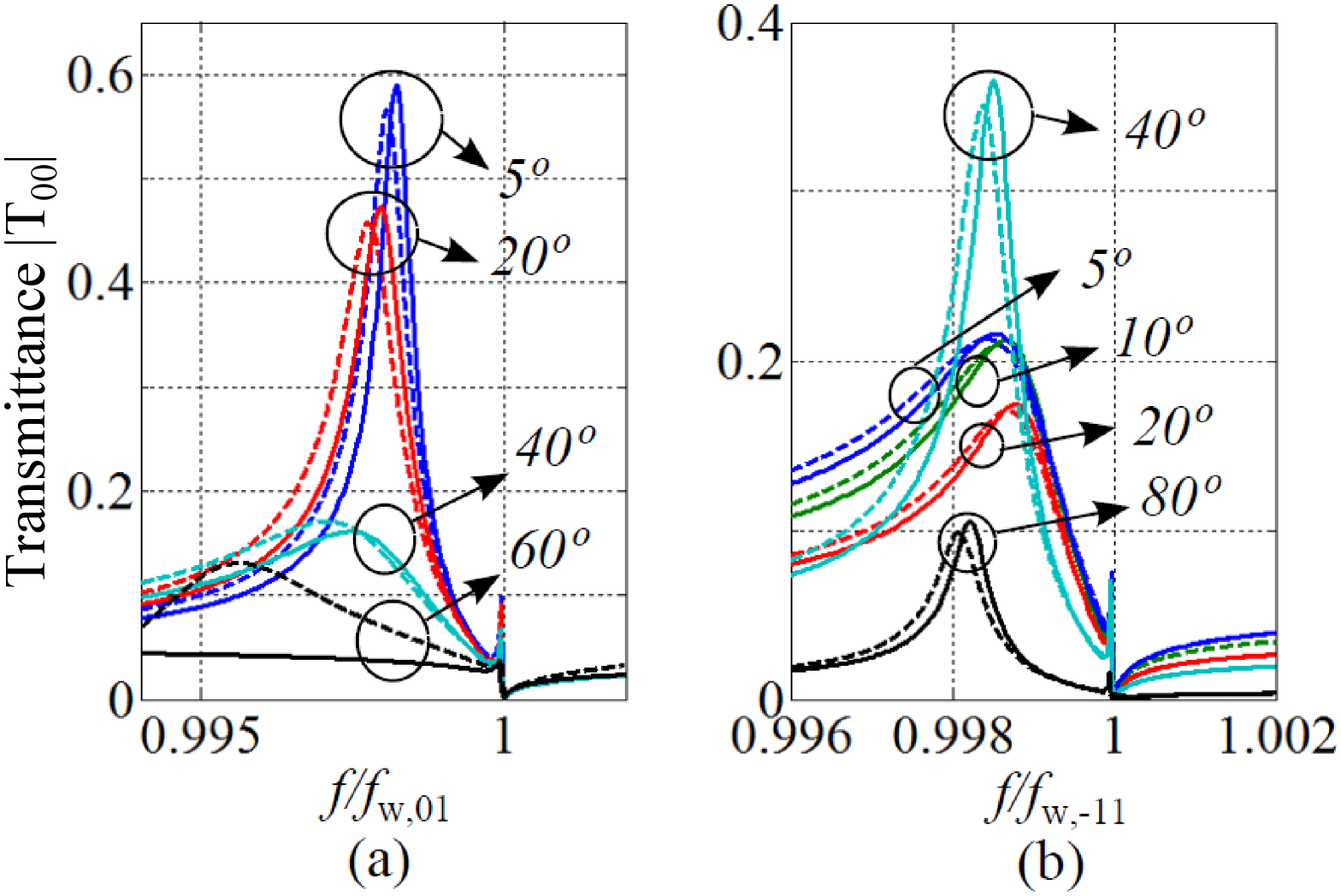}
\caption{Transmission through an array of square holes in a copper screen ($\sigma = 59.6 \times 10^6$ S/m) at oblique incidence (TE incident wave). Periodicity is $a=300{\rm{\mu}}$m, hole size is $b=a/4$, and thickness of the screen is $t=a/20$. Continuous lines correspond to the mode matching model and dashed lines to the CST simulations. In (a), the peaks correspond to the divergence of the scattered TM$_{0,1}$ mode and the Wood's frequencies range from $1.00$ THz for $\theta=5^o$ to $7.46$ THz for $\theta=60^o$. In (b), the peaks correspond to the divergence of the scattered TM$_{-1,1}$ mode and the Wood's anomaly frequencies range from $1.33$ THz for $\theta=5^o$ to $1.00$ THz for $\theta=80^o$.}
\label{Fig4}
\end{figure}

\begin{figure}[!t]
\centering
\includegraphics[height=4.6cm]{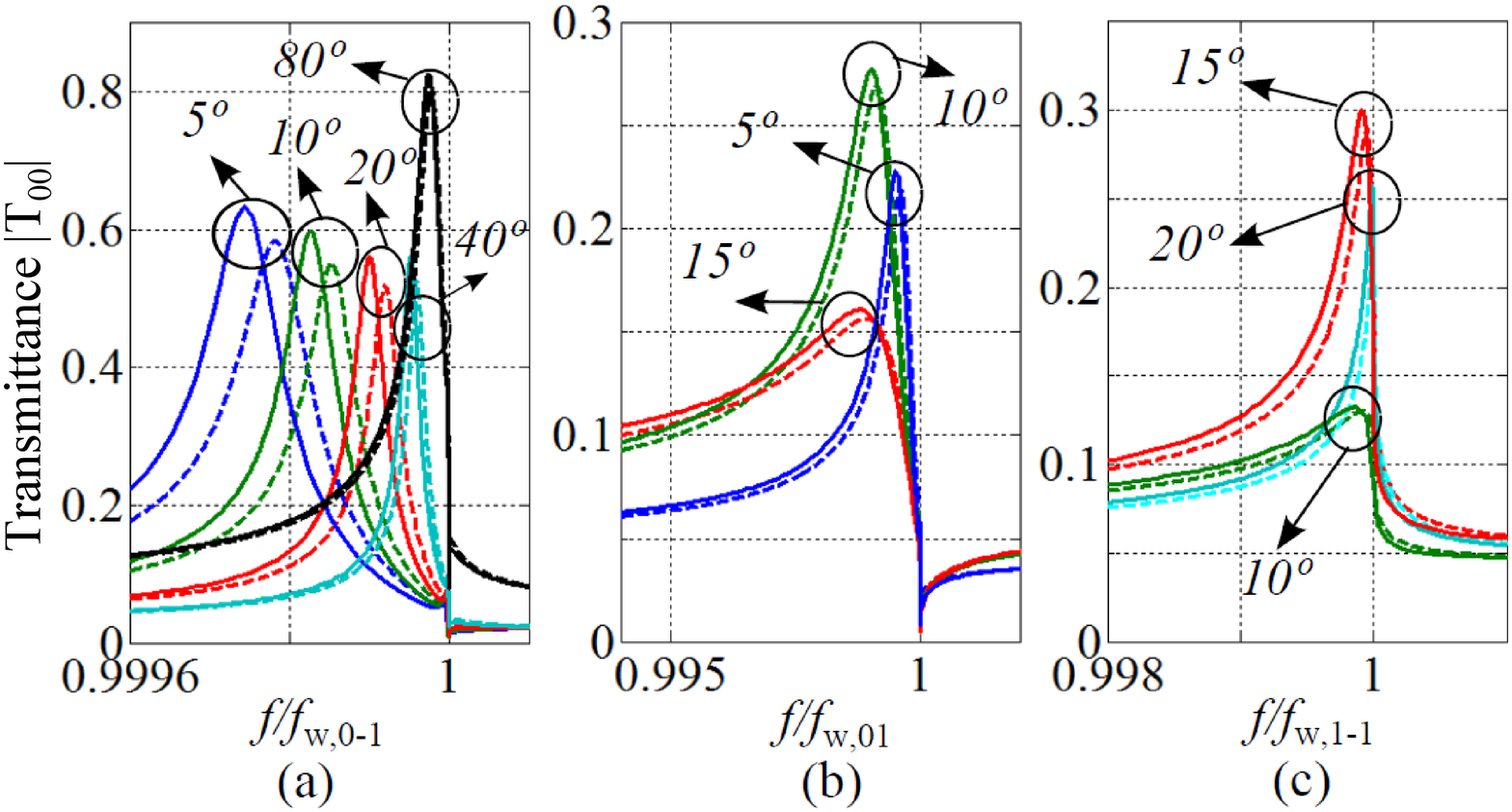}
\caption{Transmission through an array of square holes in a copper screen ($\sigma = 59.6 \times 10^6$ S/m) at oblique incidence (TM incident wave). Periodicity is $a=300{\rm{\mu}}$m, hole size is $b=a/4$ and thickness of the screen is $t=a/20$. Continuous lines correspond to the mode matching model and dashed lines to the CST simulations. In (a), the peaks correspond to the divergence of the scattered TM$_{0,-1}$ mode and the Wood's frequencies range from $0.92$ THz for $\theta=5^o$ to $0.50$ THz for $\theta=80^o$. In (b), the peaks correspond to the divergence of the scattered TM$_{0,1}$ mode and the Wood's anomaly frequencies range from $1.09$ THz for $\theta=5^o$ to $1.35$ THz for $\theta=15^o$. In (c), the peaks correspond to the divergence of the scattered TM$_{1,-1}$ mode and the Wood's anomaly frequencies range from $1.27$ THz for $\theta=10^o$ to $1.17$ THz for $\theta=20^o$.}
\label{Fig5}
\end{figure}

\begin{figure}[!t]
\centering
\includegraphics[height=4.6cm]{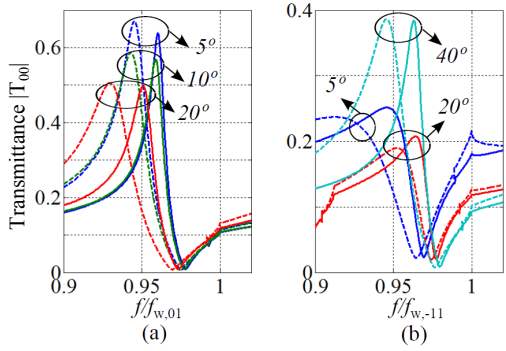}
\caption{Transmission through an array of square holes in a silver screen ($\omega_p = 2\pi\times 2175$ THz and $f_c= 4.35$ THz) at oblique incidence (TE incident wave). Periodicity is $a=1{\rm{\mu}}$m, hole size is $b=a/4$ and thickness of the screen is $t=a/20$. Continuous lines correspond to the mode matching model and dashed lines to the CST simulations. In (a), the peaks correspond to the divergence of the scattered TM$_{0,1}$ mode and the Wood's frequencies range from $300.93$ THz for $\theta=5^o$ to $318.98$ THz for $\theta=20^o$. In (b), the peaks correspond to the divergence of the scattered TM$_{-1,1}$ mode and the Wood's anomaly frequencies range from $400.07$ THz for $\theta=5^o$ to $315.08$ THz for $\theta=40^o$.}
\label{Fig6}
\end{figure}

\begin{figure}[!t]
\centering
\includegraphics[height=4.6cm]{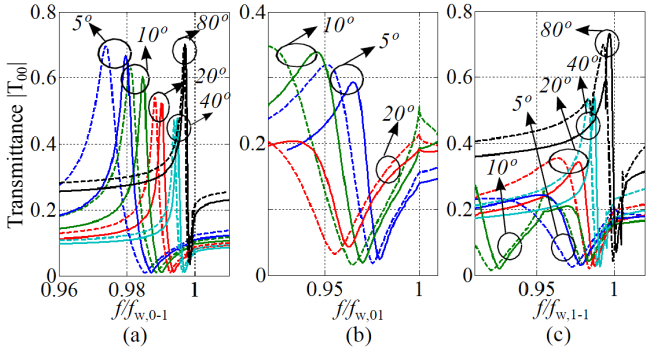}
\caption{Transmission through an array of square holes in a silver screen ($\omega_p = 2\pi\times 2175$ THz and $f_c= 4.35$ THz) at oblique incidence (TM incident wave). Periodicity is $a=1{\rm{\mu}}$m, hole size is $b=a/4$ and thickness of the screen is $t=a/20$. Continuous lines correspond to the mode matching model and dashed lines to the CST simulations. In (a), the peaks correspond to the divergence of the scattered TM$_{0,-1}$ mode and the Wood's frequencies range from $275.51$ THz for $\theta=5^o$ to $151.10$ THz for $\theta=80^o$. In (b), the peaks correspond to the divergence of the scattered TM$_{0,1}$ mode and the Wood's anomaly frequencies range from $328.39$ THz for $\theta=5^o$ to $455.68$ THz for $\theta=20^o$. In (c), the peaks correspond to the divergence of the scattered TM$_{1,-1}$ mode and the Wood's anomaly frequencies range from $400.07$ THz for $\theta=5^o$ to $299.79$ THz for $\theta=80^o$.}
\label{Fig7}
\end{figure}

In Figs. \ref{Fig2}-\ref{Fig7} the transmission coefficients obtained with the reported model are compared with electromagnetic simulations using \emph{CST Microwave Studio}. In order to obtain meaningful results, the resolution of the highest mode inside the holes ($P,Q$) must be similar to that of the input and output regions ($N,M$), i.e. $[N,M] \approx [P,Q](a/b)$. We employed $P=Q=2$,  which was enough to obtain accurate results compared with full wave electromagnetic simulations. CPU time per frequency point was $\sim 0.5$ s vs $\sim 4$ min with the electromagnetic solver. If more modes are used, computational time grows very fast (with $P^4$), making the theory unpractical. In addition, there appear numerical errors due to very small amplitudes of high evanescent modes.

In Figs. \ref{Fig2}, \ref{Fig4} and \ref{Fig5} the metallic screen is modeled by a finite conductivity $\sigma = 59.6 \times 10^6$ S/m (corresponding to copper), and the electric permittivity is given by
\begin{equation}\label{08}
\varepsilon _{\rm{met}}  \approx i\frac{\sigma }{{\omega \varepsilon _0 }}.
\end{equation}

In Figs. \ref{Fig3}, \ref{Fig6} and \ref{Fig7} the metallic screen is modeled by the Drude-Lorentz permittivity
\begin{equation}\label{09}
\varepsilon _{\rm{met}}  \approx \varepsilon _0 \left( {1 - \frac{{\omega _{\rm{p}}^2 }}{{\omega \left( {\omega  - if'_{\rm{c}} } \right)}}} \right),
\end{equation}
\noindent with the plasma frequency $\omega_{\rm p} = 2\pi\times 2175$ THz corresponding to silver \cite{Ordal-1983}. The frequency of collision corresponding to silver was corrected by the factor $(1+l_{\rm eff})/(2t)$ (where $l_{\rm eff}$ is the mean free path in bulk silver) in order to take into account the finite thickness of the screen \cite{Delgado-2010}. That is, we have chosen $f'_{\rm c}=(1+l_{\rm eff})/(2t) \times f_{\rm c, Ag} \approx 1.26\times  4.35$ THz. Moreover, in order to take into account the effect of field penetration through the walls of the holes, the size of the holes was increased to an effective width $b_{\rm{eff}}=b+2\delta$, where $\delta$ is the penetration depth of the fields inside the metal. For lossy metals at microwave or terahertz frequencies this field penetration is negligible ($\delta<<b$), however for solid plasmas it has a relevant effect. The penetration depth at optical frequencies is evaluated as $\delta=\lambda_{\rm{p}}/(2\pi)$, where $\lambda_{\rm{p}}$ is the plasma wavelength of the metal. This correction has effects on the frequency of the peaks, but has no effect on the estimation of losses. We will assume that losses inside the holes can be neglected with regard to losses in the external metal interfaces. The accuracy of this assumption is checked in the following, by comparison with electromagnetic simulations.

In the Figures, the frequencies are normalized to those corresponding to Wood's anomalies ($f_{{\rm w},nm}$) for each angle of incidence. These Wood's frequencies are defined as those frequencies at which the admittances of the different TM$_{n,m}$ modes scattered by the screen diverge. Therefore, they correspond to the solutions of the equation
\begin{equation}\label{10}
f_{{\rm w},nm}^2 (\theta )  = \left( {\frac{{nc}}{a} + f_{{\rm w},nm} \sin (\theta )} \right)^2  + \left( {\frac{{mc}}{{a}}} \right)^2
\end{equation}
\noindent for an incident TE wave, and
\begin{equation}\label{11}
f_{{\rm w},nm}^2 (\theta )  = \left( {\frac{{nc}}{a}} \right)^2  + \left( {\frac{{mc}}{{a}} + f_{{\rm w},nm} \sin (\theta )} \right)^2
\end{equation}
\noindent for an incident TM wave.

Figs. \ref{Fig2}-\ref{Fig3} correspond to the case of normal incidence and different geometries. The coupled-wave model presented here is compared with the ``small holes'' model reported in \cite{Delgado-2010}, and with the results obtained using a conventional electromagnetic solver, which were used as reference. As expected, the ``small hole'' model fails for screens with holes larger than $b=a/4$ (see Fig. \ref{Fig2}). The proposed new model provides good results even for holes as large as $b=a/2$. For larger holes it starts to be inaccurate. However, in those cases the transmission peak is due to the resonance of the hole and can not be properly called ``extraordinary''. As was mentioned earlier, these results were obtained with computation times several order of magnitude smaller than those obtained using conventional electromagnetic solvers.

In the remaining figures - Figs. \ref{Fig4} to \ref{Fig7} - oblique incidence of TE and TM waves over copper and silver screens operating at THz and optical frequencies, respectively, is considered. A quite good agreement is found between the reported model and the simulations for any angle of incidence, periodicity and hole size. Only in cases of very high angles of incidence and frequencies high enough to provide relatively small values of permittivity ($|\varepsilon _{\rm{met}}| \sim 1$ or $10$), the results of the model deviate significantly from those of the simulations. These cases, however, are of small practical interest, because the background transmission is high and the height of the peaks is small.


Since the Wood's frequency values (\ref{10}-\ref{11}) are different for the different angles of incidence, and the frequencies are normalized to Wood's values in the figures, the actual frequencies corresponding to the peaks shown in these figures cover a quite wide range of values (see captions). For an incident TM wave impinging at an angle $\theta=11.54^o$ on the screen, the Wood's frequency corresponding to the divergence of the scattered TM$_{0,1}$ and TM$_{1,-1}$ modes is the same. For higher angles of incidence, the peaks corresponding to the divergence of the TM$_{0,1}$ mode become weaker (Figs. \ref{Fig5}(b) and \ref{Fig7}(b)), and those corresponding to the divergence of the TM$_{1,-1}$ become stronger (Figs. \ref{Fig5}(c) and \ref{Fig7}(c)). A similar effect occurs for incident TE waves at $\theta=26.57^o$ and TM$_{0,1}$ and TM$_{-1,1}$ modes.

\begin{figure}[!t]
\centering
\includegraphics[height=4.2cm]{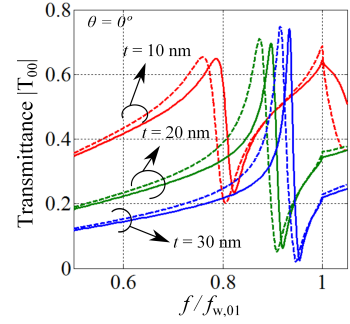}
\caption{Transmission through an array of square holes in a silver screen ($\omega_p = 2\pi\times 2175$ THz and $f_c= 4.35$ THz) at normal incidence and different thicknesses of the screen. Periodicity is $a=1{\rm{\mu}}$m and hole size is $b=a/4$. Wood's anomaly frequency is $f_{{\rm w},01}\approx 299.79$ THz. Continuous lines correspond to the mode matching model and dashed lines to the CST simulations.}
\label{Fig8}
\end{figure}

The surface impedance condition (\ref{01}) used along this analysis is also properly taking into account the tunneling through the metal, when the screen thickness becomes of the same order or smaller than the penetration depth. This effect is studied in Fig. \ref{Fig8} where EOT through silver screens of different thicknesses is studied. As expected, tunneling through the metal results in increased background transmission and smoother transmission peaks. The agreement of our results with the electromagnetic simulations is very good, with no loss of accuracy for thinner screens.

\begin{figure}[!t]
\centering
\includegraphics[height=4.2cm]{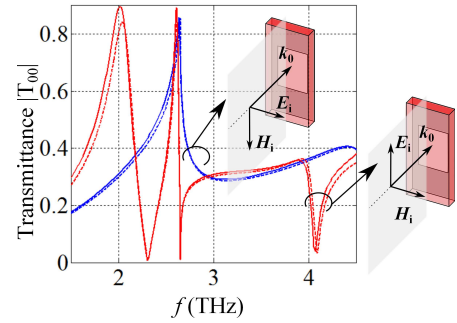}
\caption{Transmission through an array of square holes in an aluminium layer ($\omega_p = 2\pi\times 3570$ THz and $f_c= 54.11$ THz) deposited on a dielectric slab ($\varepsilon_{\rm{d}}=2.25$ and $\tan\delta=0.001$). The incident plane wave is a TEM wave with the electric field polarized either along the short (the so called anomalous extraordinary transmission \cite{Kuznetsov-2009}) and long (the so called regular extraordinary transmission \cite{Kuznetsov-2009}) periodicities. Periodicities are $47.5 \times 113{\rm{\mu}}$m, hole size is $35.2{\rm{\mu}}$m, thickness of the metallic layer is $0.5{\rm{\mu}}$m and thickness of the dielectric slab is $20{\rm{\mu}}$m. Continuous lines correspond to the mode matching model and dashed lines to the CST simulations.}
\label{Fig9}
\end{figure}

\subsection{Dielectric loaded screens} \label{Dielectric}

Realistic EOT screens must be supported in most cases by a dielectric board. The presence of the dielectrics also introduces new effects such as anomalous EOT \cite{Kuznetsov-2009}. The analysis presented in Section \ref{Theory} can be easily extended to these cases using the standard scattering matrix method. This implies to consider the presence of a complete set of electromagnetic modes impinging from $z\rightarrow\pm\infty$ in Fig. \ref{Fig1} and to compute the complete scattering matrix between the interfaces at $z=\pm t/2$ for all these modes. Once this matrix is computed, we only need to combine this matrix with the propagation matrices through the dielectric board(s), and the scattering matrices of the air-dielectric interface(s). This procedure is cumbersome but straightforward, therefore, it will not be shown explicitly. At this point it is worth noting that the cascading of scattering matrixes is not as simple as the cascading of transmission matrixes \cite{Collin}, but gives important advantage of the numerical stability with respect to ill-conditioned transmission matrixes which use both, the negative and the positive exponentials.

In Fig. \ref{Fig9} EOT through a dielectric loaded metallic screen similar to that studied in \cite{Kuznetsov-2009} (Fig. 3.a) is analyzed. In particular, we used the same periodicity and substitute circular holes by square holes of the same cross section, obtaining almost identical results, with the anomalous and ``regular'' EOT peaks clearly present.
\begin{figure}[!t]
\centering
\includegraphics[height=4.2cm]{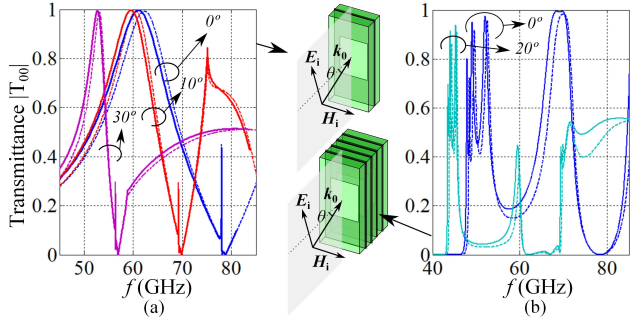}
\caption{Transmission through a single (a) and four stacked (b) copper screens ($\sigma = 59.6 \times 10^6$ S/m) at oblique incidence (TM incident wave). In (a) the metallic screen is sandwiched between two dielectric slabs ($\varepsilon_{\rm{d}}=2.43$) and in (b) the four metallic screens are placed among a total of five dielectric slabs. Periodicities are $1.5 \times 3.4$ mm, size of the square holes is $1.1$ mm, thickness of the metallic layer is $35{\rm{\mu}}$m and thickness of each of the dielectric slabs is $0.49$ mm. Continuous lines correspond to the mode matching model and dashed lines to the CST simulations.}
\label{Fig10}
\end{figure}

\subsection{Refraction of Gaussian beams}

As it is well known, laws of refraction in material slabs can be determined by analyzing the behavior of light beams incident at an oblique angle over the slab. For very large (positive or negative) index of refraction, the beam is refracted along the normal to the slab interface, and the output beam emerges from the opposite side of the slab with some lateral shift which depends on the slab thickness. Assuming this shift as the reference, positive or negative shifts with regard to this reference are the signature of positive or negative index of refraction. In \cite{Beruete-2009} and \cite{Beruete-2010-NJP} the behavior of electromagnetic Gaussian beams impinging at an oblique angle on single and stacked EOT screens was studied. It was found that in all cases TM beams experience a ``negative'' shift, whereas TE beams always experience positive or no shift, which was consistent with the presence of negative refraction in fishnet metamaterials \cite{Beruete-2009}.

\begin{figure}[!t]
\centering
\includegraphics[height=4.2cm]{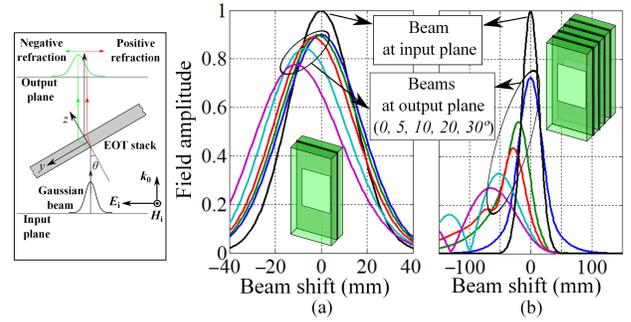}
\caption{Electric field profiles of a 1D Gaussian beam impinging at different angles of incidence and passing through the same structures as in Fig.\ref{Fig10}. For each angle of incidence, the fields are calculated at the frequency of the maximum transmission. For the single structure (a) these frequencies are $61.2$, $60.7$, $59.6$, $56.3$ and $52.7$ THz for angles of incidence $0$, $5$, $10$, $20$ and $30^0$ respectively. For the four-stacked-screens structure (b) these frequencies are $52.1$, $51.1$, $49.2$, $45.3$ and $42.1$ THz for angles of incidence $0$, $5$, $10$, $20$ and $30^0$ respectively. Deviation of the beam to the left is higher for increasing angles of incidence. The left inset shows the polarization of the beam relative to the screen(s) as well as the planes where the field profiles are depicted. The input and output planes are placed $100$ mm away from the origin of the cartesian axes.}
\label{Fig11}
\end{figure}

The scattering matrix scheme in Section \ref{Dielectric} can be also applied to stacks of EOT screens. Once the transmission coefficients for incident plane waves are known, it is a relatively straightforward task to analyze the behavior of incident Gaussian beams, which can be decomposed into plane waves. In Fig. \ref{Fig10} we show the transmission coefficient for TM plane waves impinging at several angles over dielectric loaded single and stacked EOT screens, with an structure very similar to the structures studied in \cite{Beruete-2009}. As can be seen in the figure, our results show a very good agreement with the electromagnetic simulations, with much smaller computation times. Once the position of the peaks have been determined, we are ready for the computation of the lateral shift of an incident Gaussian beam at the frequency of these transmission peaks. The profiles of the input and output TM beams impinging at different angles over the single and stacked structures are shown in Fig. \ref{Fig11}. Each output beam profile is centered on the aforementioned reference point, corresponding to an infinite ``effective refractive index'' for the structure. As can be seen, a negative shift is observed for all output beams, in agreement with the experimental results shown in \cite{Beruete-2009} and \cite{Beruete-2010-NJP}. Similar computations have been carried out for TE beams (not shown), confirming only small positive shift.

\section{Conclusion}
An analytical method has been provided for the analysis of extraordinary optical transmission (EOT) through realistic metallic screens with a periodic array of holes. Our method is based on the coupled-wave analysis with approximate surface impedance boundary conditions at the metallic interfaces. This method allows for a fast and accurate characterization of EOT for any angles of incidence, hole sizes, thicknesses and all frequencies where EOT has been reported (from microwaves to optics). Due to its computational efficiency, the method can be applied to the analysis of dielectric loaded and stacked EOT screens, including complicated sources, such as Gaussian beams. The method has been tested by comparing the results with those coming from commercial general purpose electromagnetic solvers (with computation times several orders of magnitude smaller) and experimental work on anomalous EOT and negative shift of Gaussian beams through single and stacked EOT screens. A very good agreement has been obtained in all cases. Future work will be aimed to the complete characterization of these and other exciting phenomena in single and stacked EOT screens and fishnet metamaterials.


%

\appendices
\section{}
The transversal components of the electromagnetic fields at both sides of the screen (regions 1 and 3) and inside the hole (region 2) can be expanded in terms of Bloch and waveguide modes and evaluated in the input and output surfaces of the screen.

In case of an incident TE wave with the electric field along the $y$ direction, the magnetic field along $x$ and $z$ and the propagation vector along $x$ and $z$, the transversal electric field components in the input and output surfaces of the screen are

\begin{equation}\label{12}
\begin{array}{l}
  E_x^{(1)}   =  - \sum\limits_{\scriptstyle n =  - N \atop
  \scriptstyle m = 1}^{N,M} {ik_{y,m}^{(1)} R_{nm}^{\rm TE} \exp ( {ik_{x,n}^{(1)} x} )\sin ( {k_{y,m}^{(1)} y} )}  \\ +\sum\limits_{\scriptstyle n =  - N \atop
  \scriptstyle m = 1}^{N,M} {ik_{x,n}^{(1)} R_{nm}^{\rm TM} \exp ( {ik_{x,n}^{(1)} x} )\sin ( {k_{y,m}^{(1)} y} )}  \\
  E_y^{(1)}  =  \sum\limits_{\scriptstyle n =  - N \atop
  \scriptstyle m = 0}^{N,M} {k_{x,n}^{(1)} R_{nm}^{\rm TE} \exp ( {ik_{x,n}^{(1)} x} )\cos ( {k_{y,m}^{(1)} y} )} \\
  +\sum\limits_{\scriptstyle n =  - N \atop
  \scriptstyle m = 1}^{N,M} {k_{y,m}^{(1)} R_{nm}^{\rm TM} \exp ( {ik_{x,n}^{(1)} x} )\cos ( {k_{y,m}^{(1)} y} )} \\
  + k_{x,0}^{(1)} \exp ( {ik_{x,0} x} )  \\
  E_x^{(3)}   =  - \sum\limits_{\scriptstyle n =  - N \atop
  \scriptstyle m = 1}^{N,M} {ik_{y,m}^{(1)} T_{nm}^{\rm TE} \exp ( {ik_{x,n}^{(1)} x} )\sin ( {k_{y,m}^{(1)} y} )}  \\
  + \sum\limits_{\scriptstyle n =  - N \atop
  \scriptstyle m = 1}^{N,M} {ik_{x,n}^{(1)} T_{nm}^{\rm TM} \exp ( {ik_{x,n}^{(1)} x} )\sin ( {k_{y,m}^{(1)} y} )}  \\
  E_y^{(3)}  = \sum\limits_{\scriptstyle n =  - N \atop
  \scriptstyle m = 0}^{N,M} {k_{x,n}^{(1)} T_{nm}^{\rm TE} \exp ( {ik_{x,n}^{(1)} x} )\cos ( {k_{y,m}^{(1)} y} )}  \\
  + \sum\limits_{\scriptstyle n =  - N \atop
  \scriptstyle m = 1}^{N,M} {k_{y,m}^{(1)} T_{nm}^{\rm TM} \exp ( {ik_{x,n}^{(1)} x} )\cos ( {k_{y,m}^{(1)} y} )}  \\
 \end{array}
\end{equation}
\noindent where $R_{nm}^{\rm TE}$, $R_{nm}^{\rm TM}$, $T_{nm}^{\rm TE}$ and $T_{nm}^{\rm TM}$ are the amplitudes of the different modes scattered in the screen, $k_{x,n}^{(1)}  = k_{x,0}  + \dfrac{{2n\pi }}{a}$ and $k_{y,m}^{(1)} = \dfrac{{2m\pi }}{a}$ are the transversal components of the wave vectors of the different modes, and $k_{x,0}= k_0 \rm{sin} (\theta)$ with $\theta$ being the angle of incidence. The transversal electric fields inside the holes in this case are
\begin{equation}\label{13}
\begin{array}{l}
 E_x^{(2)}  =  - \sum\limits_{\scriptstyle p = 0 \atop
  \scriptstyle q = 1}^{P,Q} {k_{y,q}^{(2)}} \Big[ {S_{pq}^{\rm TE + } \exp ( { - ik_{z,pq}^{( 2 )} z} )}  \\
  + S_{pq}^{\rm TE - } \exp ( { - ik_{z,pq}^{( 2 )} [ {z - t} ]} ) \Big] \cdot \cos ( {k_{x,p}^{(2)} (x + b/2)} )\sin ( {k_{y,q}^{(2)} y} ) \\
  \hspace{1.1cm}+ \sum\limits_{\scriptstyle p = 1 \atop
  \scriptstyle q = 1}^{P,Q} {k_{x,p}^{(2)}} \Big[ {S_{pq}^{\rm TM + }} \exp ( { - ik_{z,pq}^{( 2 )} z} ) \\
  + S_{pq}^{\rm TM - } \exp ( { - ik_{z,pq}^{( 2 )} [ {z - t} ]} ) \Big] \cdot \cos ( {k_{x,p}^{(2)} (x + b/2)} )\sin ( {k_{y,q}^{(2)} y} ) \\
 E_y^{(2)}  = \sum\limits_{\scriptstyle p = 1 \atop
  \scriptstyle q = 0}^{P,Q} {k_{x,p}^{(2)}} \Big[ {S_{pq}^{\rm TE + }} \exp ( { - ik_{z,pq}^{( 2 )} z} ) \\
  + S_{pq}^{\rm TE - } \exp ( { - ik_{z,pq}^{( 2 )} [ {z - t} ]} ) \Big] \cdot \sin ( {k_{x,p}^{(2)} (x + b/2)} )\cos ( {k_{y,q}^{(2)} y} ) \\
  \hspace{1.1cm}+ \sum\limits_{\scriptstyle p = 1 \atop
  \scriptstyle q = 1}^{P,Q} {k_{y,q}^{(2)} } \Big[ {S_{pq}^{\rm TM + }} \exp ( { - ik_{z,pq}^{( 2 )} z} ) \\
  + S_{pq}^{\rm TM - } \exp ( { - ik_{z,pq}^{( 2 )} [ {z - t} ]} ) \Big] \cdot \sin ( {k_{x,p}^{(2)} (x + b/2)} )\cos ( {k_{y,q}^{(2)} y} ) \\
 \end{array}
\end{equation}
\noindent where $S_{pq}^{\rm TE \pm }$ and $S_{pq}^{\rm TM \pm}$ are the amplitudes of the different modes inside the holes propagating either to the left ($^-$) or to the right ($^+$), and $k_{x,p}^{(2)}  = \dfrac{{p\pi }}{b}$, $k_{y,q}^{(2)}  = \dfrac{{2q\pi }}{b}$ and $k_{z,pq}^{(2)}  = \sqrt{\varepsilon _{\rm h} k_0^2  - \left( {k_{x,p}^{(2)} } \right)^2  - \left( {k_{y,q}^{(2)} } \right)^2 } $ are the components of the wave vectors of the different modes. $\varepsilon _{\rm h}$ is the relative permittivity of the material filling the holes. In all the simulations in this paper $\varepsilon _{\rm h} = 1$. Note that for this particular polarization there are virtual electric walls in $y=0,\pm a/2$.

In case of an incident TM wave with the electric field along the $y$ and $z$ direction, the magnetic field along $x$ and the propagation vector along $y$ and $z$, the transversal electric field components in the input and output surfaces of the screen are
\begin{equation}\label{14}
\begin{array}{l}
  E_x^{(1)}   =  - \sum\limits_{\scriptstyle n = 1 \atop
  \scriptstyle m =  - M}^{N,M} {ik_{y,m}^{(1)} R_{nm}^{\rm TE} \sin ( {k_{x,n}^{(1)} x} )\exp ( {ik_{y,m}^{(1)} y} )} \\
  + \sum\limits_{\scriptstyle n = 1 \atop
  \scriptstyle m =  - M}^{N,M} {ik_{x,n}^{(1)} R_{nm}^{\rm TM} \sin ( {k_{x,n}^{(1)} x} )\exp ( {ik_{y,m}^{(1)} y} )}  \\
  E_y^{(1)}   =  \sum\limits_{\scriptstyle n = 1 \atop
  \scriptstyle m =  - M}^{N,M} {k_{x,n}^{(1)} R_{nm}^{\rm TE} \cos ( {k_{x,n}^{(1)} x} )\exp ( {ik_{y,m}^{(1)} y} )}  \\
  + \sum\limits_{\scriptstyle n = 0 \atop
  \scriptstyle m =  - M}^{N,M} {k_{y,m}^{(1)} R_{nm}^{\rm TM} \cos ( {k_{x,n}^{(1)} x} )\exp ( {ik_{y,m}^{(1)} y} )} \\
  +k_{y,0}^{(1)} \exp ( {ik_{y,0} y} )   \\
  E_x^{(3)}   =  - \sum\limits_{\scriptstyle n = 1 \atop
  \scriptstyle m =  - M}^{N,M} {ik_{y,m}^{(1)} T_{nm}^{\rm TE} \sin ( {k_{x,n}^{(1)} x} )\exp ( {ik_{y,m}^{(1)} y} )}  \\
  + \sum\limits_{\scriptstyle n = 1 \atop
  \scriptstyle m =  - M}^{N,M} {ik_{x,n}^{(1)} T_{nm}^{\rm TM} \sin ( {k_{x,n}^{(1)} x} )\exp ( {ik_{y,m}^{(1)} y} )}  \\
  E_y^{(3)}   = \sum\limits_{\scriptstyle n = 1 \atop
  \scriptstyle m =  - M}^{N,M} {k_{x,n}^{(1)} T_{nm}^{\rm TE} \cos ( {k_{x,n}^{(1)} x} )\exp ( {ik_{y,m}^{(1)} y} )}  \\ + \sum\limits_{\scriptstyle n = 0 \atop
  \scriptstyle m =  - M}^{N,M} {k_{y,m}^{(1)} T_{nm}^{\rm TM} \cos ( {k_{x,n}^{(1)} x} )\exp ( {ik_{y,m}^{(1)} y} )}  \\
\end{array}
\end{equation}
\noindent where $k_{x,n}^{(1)}  = \dfrac{{2n\pi }}{a}$ and $k_{y,m}^{(1)} =  k_{y,0}  +  \dfrac{{2m\pi }}{a}$ are the transversal components of the wave vectors of the different modes scattered in the screen and $k_{y,0}= k_0 \rm{sin} (\theta)$, with $\theta$ being the angle of incidence. The transversal electric fields inside the holes in this case are
\begin{equation}\label{15}
\begin{array}{l}
 E_x^{(2)}  = \sum\limits_{\scriptstyle p = 0 \atop
  \scriptstyle q = 1}^{P,Q} {k_{y,q}^{(2)}} \Big[ {S_{pq}^{\rm TE + }} \exp ( { - ik_{z,pq}^{( 2 )} z} ) \\
   + S_{pq}^{\rm TE - } \exp ( { - ik_{z,pq}^{( 2 )} [ {z - t} ]} ) \Big]  \cdot \sin ( {k_{x,p}^{(2)} x} )\sin ( {k_{y,q}^{(2)} (y + b/2)} ) \\
  \hspace{1.1cm}- \sum\limits_{\scriptstyle p = 1 \atop
  \scriptstyle q = 1}^{P,Q} {k_{x,p}^{(2)}} \Big[ {S_{pq}^{\rm TM + }} \exp ( { - ik_{z,pq}^{( 2 )} z} ) \\
  + S_{pq}^{\rm TM - } \exp ( { - ik_{z,pq}^{( 2 )} [ {z - t} ]} ) \Big] \cdot \sin ( {k_{x,p}^{(2)} x} )\sin ( {k_{y,q}^{(2)} (y + b/2)} ) \\
 E_y^{(2)}  = \sum\limits_{\scriptstyle p = 1 \atop
  \scriptstyle q = 0}^{P,Q} {k_{x,p}^{(2)}} \Big[ {S_{pq}^{\rm TE + }} \exp ( { - ik_{z,pq}^{( 2 )} z} ) \\
  + S_{pq}^{\rm TE - } \exp ( { - ik_{z,pq}^{( 2 )} [ {z - t} ]} ) \Big] \cdot \cos ( {k_{x,p}^{(2)} x} )\cos ( {k_{y,q}^{(2)} (y + b/2)} ) \\
  \hspace{1.1cm}+ \sum\limits_{\scriptstyle p = 1 \atop
  \scriptstyle q = 1}^{P,Q} {k_{y,q}^{(2)} } \Big[ {S_{pq}^{\rm TM + }} \exp ( { - ik_{z,pq}^{( 2 )} z} ) \\
  + S_{pq}^{\rm TM - } \exp ( { - ik_{z,pq}^{( 2 )} [ {z - t} ]} ) \Big] \cdot \cos ( {k_{x,p}^{(2)} x} )\cos ( {k_{y,q}^{(2)} (y + b/2)} ) \\
 \end{array}
\end{equation}
\noindent where $k_{x,p}^{(2)}  = \dfrac{{(2p - 1)\pi }}{b}$, $k_{y,q}^{(2)}  = \dfrac{{q\pi }}{b}$ and $k_{z,pq}^{(2)}  = \sqrt{\varepsilon _{\rm h} k_0^2  - \left( {k_{x,p}^{(2)} } \right)^2  - \left( {k_{y,q}^{(2)} } \right)^2 } $  are the components of the wave vectors of the different modes inside the holes. Note that for this particular polarization there are virtual magnetic walls in $x=0,\pm a/2$.

Expressions for magnetic fields $H_x^{(i)}$ are obtained by just multiplying each mode in Eqs. (\ref{12}-\ref{15}) by its corresponding wave admittance. Expansions in Eqs. (\ref{13}) and (\ref{15}) correspond to a square waveguide with PEC walls, so losses inside the holes are neglected.

Boundary conditions (\ref{06}) and (\ref{07}) can be multiplied by a convenient set of orthogonal functions and transformed in integral boundary conditions over the screen and in the region of the holes. In the cross section of the unit cell, the following integral boundary conditions are satisfied
\begin{equation}\label{16}
\begin{array}{l}
 \mathlarger{\iint_{wg}} { {\left\{ { E_y^{(3)}  +  E_y^{(1)} - Z_{s1} \left[ { H_x^{(3)}  -  H_x^{(1)}  }\right]} \right\}  f_{nm}dS} }  \\
  -  \mathlarger{\iint_{h}} { {\left\{  {E_y^{(2)} +  {E_y^{(2)} }  - Z_{s1} \left[ {\ {H_x^{(2)} }  -  {H_x^{(2)} } } \right]} \right\} f_{nm}dS} }=0  \\ \\ \\
  \mathlarger{\iint_{wg}} { {\left\{ {\ {E_y^{(3)} }  -  {E_y^{(1)} } - Z_{s2} \left[ {\ {H_x^{(3)} }  +  {H_x^{(1)} }  } \right]} \right\}  f_{nm}dS} }  \\
  - \mathlarger{\iint_{h}} { {\left\{ {\ {E_y^{(2)} }  -  {E_y^{(2)} }  - Z_{s2} \left[ {\ {H_x^{(2)} }  +  {H_x^{(2)} } } \right]} \right\}  f_{nm}dS} }=0  \\ \\ \\
 \mathlarger{\iint_{wg}} { {\left\{ {\ {E_x^{(3)} }  +  {E_x^{(1)} }  + Z_{s1} \left[ {\ {H_y^{(3)} }  - {H_y^{(1)} } } \right]} \right\}  g_{nm}dS} }  \\
  - \mathlarger{\iint_{h}} { {\left\{ {\ {E_x^{(2)} }  +  {E_x^{(2)} }  + Z_{s1} \left[ {\ {H_y^{(2)} }  -  {H_y^{(2)} }  } \right]} \right\}  g_{nm}dS} }=0  \\ \\ \\
 \mathlarger{\iint_{wg}} { {\left\{ {\ {E_x^{(3)} }  -  {E_x^{(1)} }  + Z_{s2} \left[ {\ {H_y^{(3)} }  + {H_y^{(1)} } } \right]} \right\}  g_{nm}dS} } \\
  - \mathlarger{\iint_{h}} { {\left\{ {\ {E_x^{(2)} }  -  {E_x^{(2)} }  + Z_{s2} \left[ {\ {H_y^{(2)} }   +  {H_y^{(2)} } } \right]} \right\}  g_{nm}dS} }=0,  \\
 \end{array}       
\end{equation}
\noindent where subindex $wg$ and $h$ stand for the waveguide and hole sections respectively, and $f_{nm}$ and $g_{nm}$ are orthogonal functions in the area of the unit cell. For incident TE waves with the electric field along $y$
\begin{equation}\label{17}
\begin{array}{l}
  f_{nm} (x,y) = \exp ( {-ik_{x,n}^{(1)} x} )\cos ( {k_{y,m}^{(1)} y} ),\\ [1.2em]
 g_{nm} (x,y) = \exp ( {-ik_{x,n}^{(1)} x} )\sin ( {k_{y,m}^{(1)} y} ), \\ [1.2em]
 n = -N,...,0,..., N;\:\: m = 1, 2,..., M;  \\
 \end{array}
\end{equation}
and for incident TM waves with the magnetic field along $x$
\begin{equation}\label{18}
\begin{array}{l}
 f_{nm} (x,y) = \cos ( {k_{x,n}^{(1)} x} ) \exp ( {-ik_{y,m}^{(1)} y} ), \\ [1.2em]
 g_{nm} (x,y) = \sin ( {k_{x,n}^{(1)} x} ) \exp ( {-ik_{y,m}^{(1)} y} ), \\ [1.2em]
 n = 1, 2,.., N;\:\: m = -M,..., 0,..., M. \\
 \end{array}
\end{equation}
In the area of the holes, continuity of transverse electromagnetic fields is imposed with the following integral boundary conditions
\begin{equation}\label{19}
\begin{array}{l}
 \mathlarger{\iint_{h}} { { {H_x^{(1)} }   \cdot u_{pq} (x,y)dS} }  = \mathlarger{\iint_{h}} { { {H_x^{(2)} }   \cdot u_{pq} (x,y)dS} }  \\[12pt]
\mathlarger{\iint_{h}} { { {H_y^{(1)} }   \cdot w_{pq} (x,y)dS} }  = \mathlarger{\iint_{h}} { { {H_y^{(2)} }    \cdot w_{pq} (x,y)dS} }  \\[12pt]
 \mathlarger{\iint_{h}} { { {H_x^{(2)} }   \cdot u_{pq} (x,y)dS} }  = \mathlarger{\iint_{h}} { { {H_x^{(3)} }   \cdot u_{pq} (x,y)dS} }  \\[12pt]
 \mathlarger{\iint_{h}} { { {H_y^{(2)} }   \cdot w_{pq} (x,y)dS} }  = \mathlarger{\iint_{h}} { { {H_y^{(3)} }   \cdot w_{pq} (x,y)dS} }  \\
 \end{array}
\end{equation}
\noindent where $u_{pq}$ and $w_{pq}$ are orthogonal functions in the area of the unit cell. For incident TE waves with the electric field along $y$
\begin{equation}\label{20}
\begin{array}{l}
 u_{pq} (x,y) = \sin ( k_{x,p}(x + b/2) )\cos (k_{y,q}y ), \\ [1.2em]
 w_{pq} (x,y) = \cos ( k_{x,p}(x + b/2) )\sin (k_{y,q}y ), \\ [1.2em]
 \:\:\:\:\:p = 1, 2,.., P;\:\: q = 1, 2,.., Q; \\
 \end{array}
\end{equation}
and for incident TM waves with the magnetic field along $x$
\begin{equation}\label{21}
\begin{array}{l}
 u_{pq} (x,y) = \cos ( k_{x,p} x )\cos ( k_{y,q} (y + b/2) ),  \\ [1.2em]
 w_{pq} (x,y) = \cos ( k_{x,p} x )\cos ( k_{y,q} (y + b/2) ),  \\ [1.2em]
 \:\:\:\:\:p = 1, 2,.., P;\:\: q = 1, 2,.., Q. \\
 \end{array}
\end{equation}
After substitution of (\ref{12}-\ref{15}) into (\ref{16}) and (\ref{19}), these integrals are solved analytically, providing an sparse system of equations for the coefficients in (\ref{12}-\ref{15}).

\section*{Acknowledgment}
This work has been supported by the Spanish Ministerio de Ciencia e Innovacion under projects Consolider-EMET CSD2008-00066 and TEC2010-16948 (SEACAM), by the Spanish Junta de Andalucia under project TIC-06238 (METAMED), and by the Czech Technical University in Prague (project No. SGS10/271/OHK3/3T/13).

\ifCLASSOPTIONcaptionsoff
  \newpage
\fi


\begin{thebibliography}{1}
\bibitem{Note} This equation must be seen as a pure mathematical definition, with no physical meaning. Note that in the surface impedance approximation, the fields inside the metal are irrelevant and we can thus define them to have any value.
\end{thebibliography}
\end{document}